\documentclass[aps,prl,twocolumn,showpacs,preprintnumbers,amsmath,amssymb]{revtex4}
\usepackage{amsmath,amssymb}
\usepackage{graphicx}

\makeatletter
\newif\if@preliminary
\@preliminaryfalse
\def\preliminary{\@preliminarytrue}
\newcommand{\GeV}{\mathrm{GeV}}
\newcommand{\TeV}{\mathrm{TeV}}

\newcommand{\rep}[1]{\mbox{\boldmath$#1$}}%
\newcommand{\arep}[1]{\mbox{\boldmath$\overline{#1}$}}%

\newcommand{\vev}[1]{{\langle #1 \rangle}}

\begin{document}
\preprint{DESY 06--099\;\;hep-ph/0606277\\[0.5\baselineskip] 27.~June 2006}
\title{%
    Unification Without Doublet-Triplet Splitting}
\author{Wolfgang Kilian}

\email{wolfgang.kilian@desy.de}
\affiliation{%
 Fachbereich Physik, University of Siegen, 57068 Siegen, Germany}
\affiliation{%
 Deutsches Elektronen-Synchrotron DESY, 22603 Hamburg, Germany}

\author{J\"urgen Reuter}
\email{juergen.reuter@desy.de}
\affiliation{%
 Deutsches Elektronen-Synchrotron DESY, 22603 Hamburg, Germany}


\begin{abstract}%
{Matter-Higgs unification in string-inspired supersymmetric Grand
  Unified Theories predicts the existence of colored states in the
  Higgs multiplets and calls for two extra generations of Higgs-like
  fields ('unhiggses').  If these states are present near the TeV
  scale, gauge-coupling unification points to the existence of two
  distinct scales, $10^{15}\;\GeV$ where right-handed neutrinos and a
  Pati-Salam symmetry appear, and $10^{18}\;\GeV$ where complete
  unification is achieved.  Baryon-number conservation, while not
  guaranteed, can naturally emerge from an underlying flavor symmetry.
  Collider signatures and dark-matter physics may be drastically
  different from the conventional MSSM.}
\end{abstract}

\maketitle


Grand unified theories (GUT) of all particle-physics interactions have
drawn great attention since the classic path of $SU(5)$ unification
has been discovered~\cite{Georgi:1974sy}.  In the supersymmetric
version (MSSM)~\cite{SUSY-GUTs,GUT-Rev}, the running effective
couplings of the Standard-Model (SM) gauge group meet almost exactly
at the GUT scale $M_\text{GUT}\approx 10^{16}\;\GeV$.  The
$SO(10)$ extension~\cite{SO10} furthermore incorporates right-handed
neutrinos, while trinification ($SU(3)^3\otimes
Z_3$)~\cite{trinification,e6sub,Willenbrock:2003ca} and $E_6$
GUTs~\cite{Gursey:1975ki} unify Higgs and matter representations.

All mentioned models share the famous doublet-triplet splitting
problem~\cite{DTS}: embedding all states (including Higgs) in complete
representations implies the existence of a pair of 'exotic'
color-triplet electroweak-singlet superfields $D$ and $D^c$.
Higgs-matter unification furthermore introduces two extra
Higgs/$D$/$D^c$ generations.  If these have electroweak-scale masses,
their effect on the running couplings spoils unification.
GUT-invariant $D/D^c$ superpotential interactions contain both diquark
and leptoquark couplings and thus induce rapid proton
decay~\cite{protondec}.  To avoid this problem, they are usually
placed near the GUT scale, although this cannot be explained without
further structure beyond the gauge symmetry.

Relaxing GUT constraints on the superpotential, e.g., in
'string-inspired' $E_6$ models~\cite{Witten:1985xc}, the proton-decay
problem can be removed by imposing discrete symmetries that eliminate
either diquark or leptoquark couplings~\cite{Campbell:1986xd}.
However, the lack of coupling unification seems to disfavor $D/D^c$ in
the light spectrum; recent phenomenologal studies~\cite{ESSM} rely on
extra states from incomplete representations to make conventional
unification work.

Nevertheless, it is worthwhile to inspect the renormalization-group
flow of a pure spectrum of complete GUT representations.  In
particular, we consider the fundamental $\rep{27}$ of $E_6$ that
includes, for each matter generation, all quark and lepton fields,
Higgs, $D/D^c$, and a colorless singlet $S$.  

\begin{figure}
\begin{center}
\includegraphics[width=80mm]{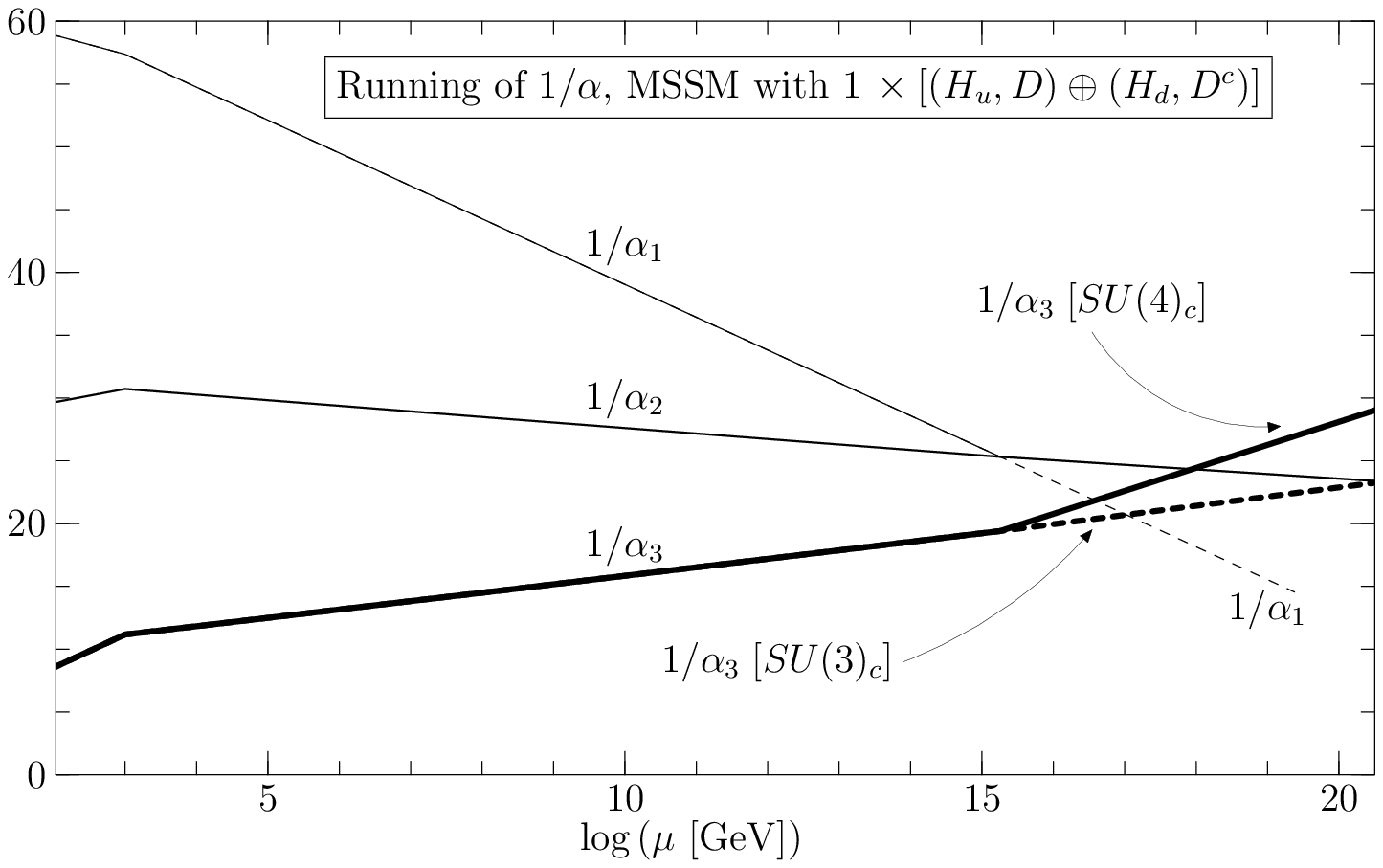}
\includegraphics[width=80mm]{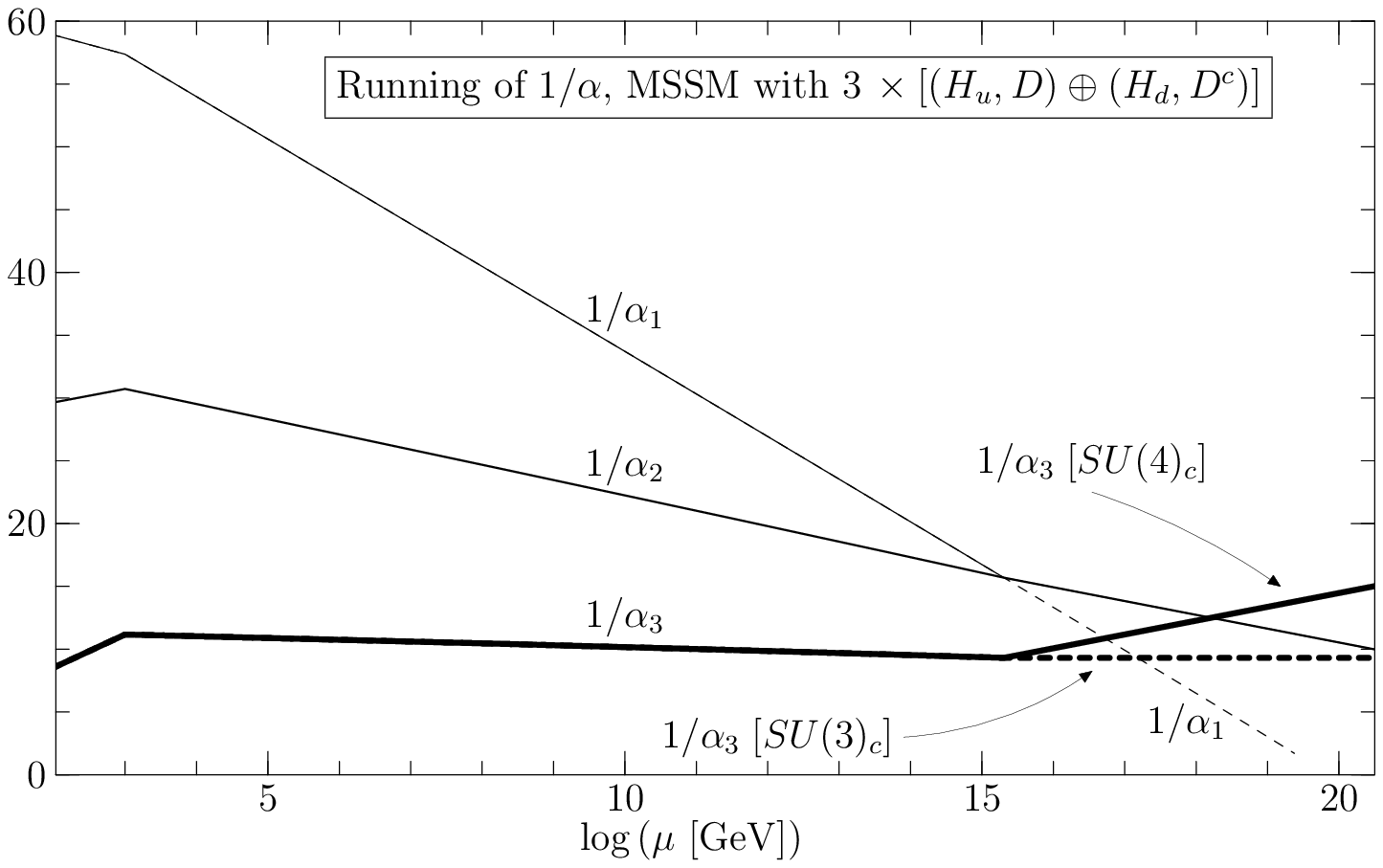}
\end{center}
\vspace{-5mm}
\caption{One-loop running couplings for the MSSM spectrum with one
  (top) or three (bottom) complete matter/Higgs families and
  left-right symmetry above the neutrino mass scale.  Dashed:
  $SU(3)_C\times SU(2)_L\times SU(2)_R$ (SM with left-right symmetry).
  Full: $SU(4)_C\times SU(2)_L\times SU(2)_R$ (Pati-Salam). The
  $SU(2)_L$ is the same in both cases.  
}
\label{fig:uni2}
\end{figure}

In such a model, the three running gauge couplings do not intersect at
a single point.  Still, the $SU(2)_L$ and $U(1)_Y$ couplings do meet
at $10^{15}\;\GeV$ (Fig.~\ref{fig:uni2}). This is a realistic mass
scale for right-handed neutrinos $\nu^c$, in accordance with see-saw
numerics if we take into account a possible hierarchy in right-handed
neutrino masses.  Adding the $\nu^c$ fields at that scale, the SM
gauge group may be extended to a left-right symmetric model.  The
combined left-right weak coupling runs differently from the QCD
coupling and unifies with it at $10^{21}\;\GeV$.  This is, however,
above the Planck scale and thus does not lead to a viable GUT model.

Instead, we can achieve unification below the Planck scale by slightly
extending this model.  At $10^{15}\;\GeV$, we postulate the appearance
of the Pati-Salam (PS) gauge symmetry $SU(4)_C\times SU(2)_L\times
SU(2)_R\times Z_2$~\cite{PS}. The corresponding renormalization-group
flow is shown in Fig.~\ref{fig:uni2} (full lines).  The ultimate
unification scale where a symmetry such as $SO(10)$ or $E_6$ could
emerge is about $10^{18}\;\GeV$, perfectly consistent with string
theory~\cite{string-rev}.  (For alternative multi-step unification
scenarios with and without exotic matter cf.~\cite{alternative}.)

Breaking the PS gauge group down to the MSSM gauge group at
$10^{15}\;\GeV$ is achieved by just integrating out the right-handed
neutrinos, so there is no need for a breaking mechanism beyond the one
that generates $\nu^c$ Majorana masses.  Furthermore, PS gauge bosons
do not mediate proton decay.  Considering gauge-superfield exchange
only, the proton-decay scale is thus lifted to $10^{18}\;\GeV$.

Regarding proton decay via superpotential interactions, the PS
symmetry does not put Higgs fields and $D/D^c$ in a common multiplet;
they fill independent $\rep{1}_{2,2}$ and $\rep{6}_{1,1}$
representations, respectively.  However, coupling unification in this
model requires their coexistence in the light spectrum.  Coupling the
latter to matter fields in a PS-invariant way simultaneously induces
leptoquark and diquark couplings and thus rapid proton decay.  Such a
term is unfortunately required to make the $D$ particles decay.

To eliminate the dangerous terms we can make use of the fact that
there are three generations of matter.  In the absence of a
superpotential, a model with three matter/Higgs families has a
$SU(3)_F$ flavor symmetry.  (The non-anomalous $SO(3)_F$ subgroup
suffices.)
Imposing flavor symmetry on the diquark couplings on top of $SU(2)_L$
and $SU(2)_R$ symmetry uniquely selects the structure
\begin{align*}
  D q_L q_L &= \epsilon^{abc}\epsilon_{\alpha\beta\gamma}\epsilon_{jk}
  D^a_{\alpha} (q_L)^b_{\beta j} (q_L)^c_{\gamma k}
\end{align*}
($D^c$ analogous), where $(a,b,c)$, $(\alpha,\beta,\gamma)$, and
$(i,j,k)$ are indices in flavor, color, and $SU(2)_L$ space,
respectively.  Due to the total antisymmetry of three $\epsilon$
symbols, this term exactly vanishes.  This property continues to hold
if we impose larger gauge symmetries (PS, $SO(10)$, $E_6$) on the
superpotential, as long as $SU(2)_L\times SU(2)_R$ is a subgroup.
Flavor symmetry in conjunction with color and left-right symmetry thus
eliminates diquark couplings, and baryon number automatically emerges
as a symmetry of the superpotential.

While $SU(2)_L$ is exact down to the $\TeV$ scale, breaking $SU(2)_R$
and $SO(3)_F$ at high energies might re-introduce $D/D^c$ diquark
couplings and thus proton-decay operators.  To exclude them, we have
to impose baryon number on all $SU(2)_R$- and $SO(3)_F$-breaking
spurions that connect $D$ or $D^c$ with other fields: they have to
involve quark fields in color-singlet pairs (or respect flavor
symmetry).  This is easy to realize since any spontaneous
symmetry-breaking can be associated to condensates at most bilinear in
fields of the fundamental representation.  After integrating out gauge
superfields, baryon number emerges as an exact symmetry of the
low-energy theory.  The flavor symmetry needs not leave obvious
traces.


We illustrate this in form of a toy model that does not refer to
specific features of supergravity or string theory.  Let us assume
that all matter and associated fields derive from fundamental
$\rep{248}$ representations of $E_8$~\cite{Matsuoka:1986vg}, which has
$E_6\times SU(3)$ as a maximal subgroup.  We identify $E_6$ with the
GUT symmetry and $SU(3)$, departing from the standard construction for
string compactification~\cite{Witten:1985xc}, with flavor.  The
multiplet decomposes into a flavor-triplet of matter $\rep{27}_3$, a
mirror image triplet $\arep{27}_{\bar 3}$, an $E_6$ adjoint
$\rep{78}_1$, and the $SU(3)_F$ adjoint $\rep{1}_8$
(cf.~e.g.~\cite{Slansky:1981yr}).


We may assume an infinite (Kaluza-Klein) tower of such
multiplets and introduce quartic couplings which in the $E_6$
decomposition contain $(\rep{27}_{3})_i(\arep{27}_{\bar
3})_i(\rep{27}_{3})_j(\arep{27}_{\bar 3})_j$.  An asymmetric spurion
(condensate) $\vev{(\rep{27}_3)_i^a(\arep{27}_{\bar
3})_j^b}=\delta^{ab}\delta_{j,i+1}$ breaks $E_8$ and removes all
mirror matter from the massless spectrum, leaving one zero mode
$(\rep{27}_3)_0$.  To reduce the symmetry further down to the
Pati-Salam group, we introduce a spurion
$\vev{\arep{1}_{2,2}\arep{1}_{2,2}}$, i.e., the $\mu$-term-type
coupling of mirror-Higgs superfields which occurs in the decomposition
of $\arep{27}_{\bar 3}\arep{27}_{\bar 3}$.  This also breaks flavor
symmetry.  To distinguish the third generation, we can also allow for
$\vev{\arep{27}_{\bar 3}}\sim\vev{\arep{1}_{1,1}}$.  (By itself, the
latter would break $E_6$ to $SO(10)$, the standard GUT path.)

Similar to the diquark coupling discussed before, the trilinear $E_6$
superpotential $\rep{27}_3\,\rep{27}_3\,\rep{27}_3$ vanishes
identically if flavor symmetry is imposed, so all matter
self-interactions are effectively generated by symmetry breaking.
Looking at other trilinear terms, we can have
$(\rep{27}_3\,\rep{78}_1\,\arep{27}_{\bar 3})$,
$(\rep{78}_1\,\rep{78}_1\,\rep{78}_1)$,
$(\rep{27}_3\,\rep{1}_8\,\arep{27}_{\bar 3})$, and
$\rep{1}_8\,\rep{1}_8\,\rep{1}_{8}$.  The effective superpotential
results from inserting condensates for $\arep{27}_{\bar 3}$ and
integrating out the remaining fields in $\rep{78}_1$ and $\rep{1}_8$.
For these, $E_6\times SU(3)_F$ invariance allows for mass terms.

This construction generates all MSSM superpotential terms, subject to
Pati-Salam symmetry, and couplings $SD^cD$ and $SH_uH_d$.  Concerning
baryon number, the only dangerous term is
$\rep{78}_1\,\rep{78}_1\,\rep{78}_1$ which after inserting the
(colorless) condensates into $\rep{27}_3\,\rep{78}_1\,\arep{27}_{\bar
3}$ and integrating out the $\rep{78}_1$ results in additional
trilinear matter couplings.  However, the color-triplet leptoquarks
$X$ contained in the $\rep{78}_1$ do not have a self-coupling: $XXX$
again vanishes by total antisymmetry with respect to all color, left,
and right indices.

A field with right-handed neutrino quantum numbers is present among
the color- and flavorless fields contained in the $\rep{78}_1$.  If
this condenses, a quartic term $(\rep{27}\,\rep{78}\,\arep{27})^2$ in
the effective superpotential generates a right-handed neutrino mass in
accordance with the seesaw mechanism~\cite{ps_neutrino} and breaks the
PS symmetry down to the SM gauge group.  Near the electroweak scale,
an $S$ condensate generates a $\mu$ term, and standard radiative
breaking of the electroweak symmetry can occur.  

We do not attempt to extend this model to a full theory of gauge and
flavor structure.  Further studies should answer the question whether
some construction that implements the main ideas can account for
realistic mixing patterns in both quark and lepton sectors.


Departing from particular toy-model features, the low-energy
particle spectrum of the Pati-Salam GUT models discussed here is a
subset of the well-known spectrum of $E_6$-type models, without the
need for Higgs states outside the $E_6$ matter
multiplets~\cite{Hewett:1988xc}.  It depends on the number of non-MSSM
(Higgs, leptoquark) generations that survive at low energies.  The
minimal version contains just the MSSM spectrum, augmented by one
color-triplet of $D$ leptoquark superfields.  The maximal version
displays three full families of $E_6$ matter including three $D$
leptoquarks, three families of superfields with the quantum numbers of
the MSSM Higgs doublets, and three SM singlet superfields~$S$.

A natural mechanism to keep the complete spectrum light is to have a
$U(1)$ subgroup of the $E_6$ gauge symmetry unbroken.  Soft
supersymmetry (SUSY) breaking would trigger radiative breaking of this
extra $U(1)$ at the $\TeV$ scale.  All Higgs and leptoquark fields
have $U(1)$ charges such that GUT-scale mass terms are excluded,
implying the existence of a $Z'$ boson in the $\TeV$
range~\cite{London:1986dk}.  Alternatively, this local symmetry may be
broken at the high scale.  In that case, vacuum expectation values
(VEVs) for $S$ fields could provide GUT-scale mass terms for some
Higgs/leptoquark families. To avoid axions, GUT-scale $U(1)$ breaking
should, via higher-dimensional terms, induce explicit $U(1)$-breaking
in the low-energy Lagrangian. These terms could be in the
superpotential ($S^3$) or in the soft SUSY-breaking part.

At the scale of soft SUSY breaking, VEVs are allowed for the neutral
components of $H_u,H_d$, and for~$S$.  In all non-minimal versions of
the model, these condensates, $\vev{H_u},\vev{H_d},\vev{S}$, are
vectors in family space.  The Higgs and $S$ superfield vectors can be
rotated such that only one component, the third one, gets a VEV and
provides MSSM-like $H_u$ and $H_d$ scalars and higgsinos.  

Yukawa couplings to matter are possible also for the two unhiggs
generations $h_u,h_d,\sigma$ that do not get a
VEV~\cite{Ellis:1985yc}.  To avoid FCNCs via double exchange of
charged unhiggses, the Yukawa matrix entries for them should either be
small~\cite{ESSM} or vanish exactly~\cite{Campbell:1986xd}.  The
latter case is equivalent to an extra $Z_2$ symmetry, $H$-parity,
which is odd just for the unhiggs superfields.  Conservation of
$H$-parity would make the lightest unhiggs (or unhiggsino) a
dark-matter candidate~\cite{Griest:1989ew}, adding to the lightest
superparticle (LSP) as the dark-matter candidate of $R$-parity
conservation.

Incidentally, baryon-number conservation via flavor symmetry
eliminates the need for $R$-parity conservation.  If we keep lepton
number as an accidental symmetry below PS-breaking, $R$-parity
conservation emerges as a derived result.  Alternatively, we could
drop lepton number and thus introduce the full phenomenology of
$R$-parity violation, while dark matter is provided by unhiggses.

Even if the unhigges $h_u,h_d$ have negligible couplings to ordinary
matter, they can still be pair-produced at colliders. Their decays
involve ordinary Higgses (including singlets), gauge bosons, or
charginos and neutralinos.  Some of these signals are detectable at
the LHC, all are easily identifiable at the ILC.  Unhiggses could also
occur in decay cascades of higher-level Higgses, charginos, and
neutralinos if kinematically allowed.  Alternatively, if $H$-parity
does not play a role, unhigges may couple significantly to some light
quarks and leptons.  In this case, there is resonant production in
$q\bar q$ annihilation. 

The particles associated with singlet superfields $S$ consist of one
scalar, one pseudoscalar, and one neutralino each.  They are all
neutral and mix with other Higgs and higgsino states.  Production and
decay occurs via mixing only, signals are thus similar to MSSM Higgses
and neutralinos.

The leptoquark superfields $D$ and $D^c$ acquire Dirac masses
proportional to $\vev{S}$.  The masses are considerably enhanced by
renormalization-group running, but some of them could be suppressed by
small Yukawa couplings to~$S$.  Thus, at the LHC we expect up to three
down-type scalar leptoquarks with arbitrary masses.  Depending on the
structure of leptoquark couplings, various decay patterns are
possible.  The most likely variant is dominant coupling to the third
generation, so leptoquarks are pair-produced in $gg$ fusion and decay
into $t\tau$ or $b\nu_\tau$ final states.  They would also show up in
cascades of gluino or squark decay, if kinematically allowed.  The
superpartners (`leptoquarkinos') should be somewhat lighter, decaying
into $\tilde t\tau$, $t\tilde\tau$, $\tilde b\nu_\tau$, or
$b\tilde\nu_\tau$.

The role of flavor symmetry in prohibiting diquark couplings of $D$
fields suggests another scenario: if the dominant terms that induce
leptoquark couplings exhibit flavor symmetry, leptoquark decays
involve all generations simultaneously. This would lead to distinctive
signatures such as $t\mu$, $te$, or light jet plus $\mu$ or $\tau$.
Additional production channels $gq\to D\ell$ would appear.  Analogous
statements hold for the corresponding fermion superpartners~$\tilde D$.


In conclusion, we have explored a SUSY-GUT scenario without
doublet-triplet splitting, such that the low-energy spectrum contains
color-triplet leptoquarks~$D$ and their superpartners.  The constraint
of gauge-coupling unification then points to the existence of two
distinct high-energy scales.  The first threshold is at
$10^{15}\;\GeV$ where the MSSM gauge group is extended to the
Pati-Salam group $SU(4)_C\times SU(2)_L\times SU(2)_R$ and
right-handed neutrino masses are generated.  At the higher energy
$10^{18}\;\GeV$, slightly below the Planck scale, complete unification
(e.g., $E_6$) is located, possibly in the context of a superstring
theory.  

While gauge interactions in a Pati-Salam GUT do not trigger proton
decay, proton decay via superpotential terms can be eliminated by an
underlying flavor symmetry.  $R$-parity conservation is no longer a
requirement.  Combining this with another symmetry that forbids FCNCs,
sources for dark matter different from the MSSM appear.  Since the
model has a considerable quantity of new parameters in the
superpotential, there are no unique predictions for their masses or
non-gauge interactions.  The LHC will certainly allow for discovery or
exclusion of leptoquark/-inos up into the $\TeV$ range.  The
(optional) presence of a $Z'$ could easily be established.  On the
other hand, some weakly interacting states are easy to miss at the
LHC, and a thorough analysis at the ILC will likely be needed for
completely uncovering this sector.

\subsection*{Acknowledgements}
We are grateful to W.~Buchm\"uller, S.F.~King, and P.M.~Zerwas for
useful comments and discussions. This work has been supported by the
Deutsche Helmholtz-Gemeinschaft, Grant No.\ VH--NG--005.


\end{document}